# A Data-Driven Approach to Coarse-Graining Simple Liquids in Confinement

Ishan Nadkarni[1]†, Haiyi Wu[2]† and Narayana. R. Aluru[1,2]*

1. Walker Department of Mechanical Engineering, The University of Texas at Austin, Austin 78712, United States
2. Oden Institute for Computational Engineering and Sciences, The University of Texas at Austin, Austin 78712, United States

†These authors contributed equally to this work
* Correspondence to aluru@utexas.edu (N.R.A.)

**Abstract:** We propose a data-driven framework for identifying coarse-grained (CG) Lennard-Jones (LJ) potential parameters in confined systems for simple liquids. Our approach involves the use of a Deep Neural Network (DNN) that is trained to approximate the solution of the Inverse Liquid State (ILST) problem for confined systems. The DNN model inherently incorporates essential physical characteristics specific to confined fluids, enabling accurate prediction of inhomogeneity effects. By utilizing transfer learning, we predict single-site LJ potentials of simple multiatomic liquids confined in a slit-like channel, which effectively replicate both the fluid structure and molecular force of the target All-Atom (AA) system when the electrostatic interactions are not dominant. In addition, we showcase the synergy between the data-driven approach and the well-known Bottom-Up coarse-graining method utilizing Relative-Entropy (RE) Minimization. Through sequential utilization of these two methods, the robustness of the iterative RE method is significantly augmented, leading to a remarkable enhancement in convergence.

## 1. Introduction

Coarse-graining has proven to be a crucial tool in modeling complex molecular systems and studying their behavior at larger spatiotemporal scales that would be impossible with high-resolution methods like All-Atom Molecular Dynamics or first-principle techniques like Ab-Initio Molecular Dynamics (AIMD).[1-2] As an example, in recent research,[3] the utilization of coarse-grained models allowed for the examination of molecular transport in nuclear pore complexes over timescales of tens of milliseconds. This remarkable feat involved systems containing approximately 200 million atoms, a task that would have been inconceivable using alternative techniques.

By reducing the degrees of freedom of a fine-grained atomic system, coarse-graining allows for a close replication of its properties at a much lower computational cost. This is analogous to solving the "Inverse Problem of Liquid State Theory" in classical mechanics,[4-8] which involves finding interatomic potential parameters that correspond to a given equilibrium particle distribution. The distribution represents a high-resolution atomistic system, while the derived potentials are for a low-resolution coarse-grained system. In the case of pairwise interactions between particles, the Bogolyubov–Born–Green–Kirkwood Yvon (BBGKY)[4] hierarchy provides a set of equations that relate the $n$ particle and $n+1$ particle distribution functions to the underlying interaction potential. Nonetheless, these equations cannot be readily applied as the higher-order distribution functions remain unknown. Even though certain approximation methods[9] have been used to estimate solutions, their applicability is limited to specific categories of materials. This highlights that finding a general solution to the inverse liquid state problem is a non-trivial task.

Several techniques for coarse-graining have been developed in the past for studying systems like[64-68] polymers, liquid crystals, colloids and biomolecules like proteins, lipid bilayers, etc. These techniques can be broadly classified into two categories: Top-Down and Bottom-Up.

Top-Down[10] approaches parameterize CG potentials by fitting them to experimental data to match thermodynamic quantities or macroscale properties like density, pressure, or surface tension[11-12]. On the other hand, Bottom-Up[13] approaches rely on principles of statistical mechanics, which govern atomic motion at molecular scales. A central idea in Bottom-Up method is to model the many-body Potential of Mean-Force (PMF) which contains necessary information required to infer key properties of the AA system. However, they rely on data from more accurate atomistic simulation, which are computationally expensive and restricted to smaller systems. The most popular Bottom-Up methods include, Iterative Boltzmann Inversion (IBI),[14] Relative Entropy (RE)[15, 62, 63] and Force-Matching (FM).[16-18] These Bottom-Up methods have certain limitations such as a deficiency in robustness when it comes to identifying a global minimum and sensitivity to initial conditions. Furthermore, potentials derived using Bottom-Up methods exhibit substantial deviations on target properties[69-70] away from the reference state-point for which they are derived and vary significantly with temperature[71-74] and density[75, 81]. Consequently, achieving transferability to different thermodynamic states has proven to be quite challenging[76-78]. In the case of iterative methods like IBI and RE, multiple MD simulation of sufficient duration are required to be run to obtain an adequate sampling of snapshots to compute quantities required for update- calculations. The originally proposed IBI[14] and RE[15,63] schemes require running simulations at each step. However, developments utilizing trajectory reweighing techniques based on statistical perturbation theory have been suggested for RE[79] and quite recently for IBI[80]. This approach eliminates the need to run a CG simulation at each iterative step. Nonetheless, it's important to note that convergence cannot be assured without careful treatment of the iterative process, in addition to the computational complexity involved. Furthermore, techniques such as IBI entail the computation of the atom distribution through the utilization of the Radial Distribution Function (RDF), which is specifically defined for homogeneous systems. This limitation restricts its applicability to confined systems, such as nanochannels where the fluid structure changes along the width.

With recent advances in ML, researchers have sought to address many of these drawbacks by using data-driven approaches.[19-29] ML based methods are fast and robust, and exciting breakthroughs have been made using them.[30] DeepILST[23] was one of the first approaches to provide an approximate solution to the inverse problem for a wide class of simple liquids[19] using a data-driven method. It used a DNN which was trained to learn the mapping between a given RDF and the corresponding potential parameters for LJ liquids at a given thermodynamic state. Transfer learning was then used to coarse-grain different multiatomic simple liquids. While this framework provides a route to obtaining a one-shot solution to the inverse problem, its applicability is limited to homogeneous systems. Numerous engineering systems, particularly those involving fluids in confined spaces such as nanochannels and interfaces, possess distinct characteristics that set them apart from their bulk counterparts. These confinement effects give rise to unique properties that have been harnessed for various applications,[31-35] including drug delivery systems, energy storage devices, and sensors. Consequently, there is significant interest in exploring the inverse problem for an inhomogeneous system and subsequently developing a tailored coarse-graining framework for confined fluids. We address this in our current work and develop a data-driven approach to coarse-grain simple liquids in a nanochannel by explicitly incorporating key features that describe confinement effects.

These effects give rise to strong density oscillations near the wall-fluid interface, resulting from the intricate interplay between attractive and repulsive forces between the wall and the fluid atoms. A strong repulsion by the wall leads to a zero-density value close to the wall. As we move away from the wall, at a short distance a fluid density peak is observed predominantly due to attractive interactions of the wall and its location is almost entirely determined by the attractive minimum of the wall fluid-potential.[36] This peak is followed by a density-minima caused due to repulsive forces against the fluid atoms of the first layer. At a distance far away from the wall the distribution of fluid particles is no longer influenced by the wall and resembles a homogeneous bulk-like structure. This coupling between wall-fluid and fluid-fluid interaction forces is therefore critical in determining the structure of confined fluids. Using theoretical arguments, we show how both force components contain necessary information for obtaining coarse-grained wall-fluid and fluid-fluid potential parameters. We then use a DNN which learns this non-trivial relationship between the LJ potential parameters and the wall-fluid/fluid-fluid force profiles of LJ liquids, providing an "approximate solution" to the Inverse Problem of Liquid State for inhomogeneous systems and subsequently a route to coarse-grain simple liquids in confinement.

We regard the obtained solution as an approximation primarily due to the modeling of the interaction potential, which is simplified as a pairwise Lennard-Jones potential. Despite its simplicity, it is one of the most frequently used potentials in molecular dynamics due to its physical origins and ability to describe essential aspects of atomic and molecular interactions, including Pauli's repulsion at close distances and London Dispersion forces at longer distances. As a result, it has been widely used to model diverse systems including real simple liquids[37] and to validate several classical theories in Statistical Mechanics which rely on assumption of pairwise interactions.[4] Although its use also imposes constraints in terms of applicability to complex molecules, (which have strong multi-body correlation or long-range interactions which can't be captured by the LJ functional form[38]) it also retains an element of interpretability because of its physically derived basis. We seek to alleviate this limitation and broaden its applicability to complex molecules by using a sequential approach that combines the data-driven method with Relative Entropy Minimization-based coarse-graining method. The RE derived CG potentials have been successfully used to coarse-grain complex liquids (like water) in confinement[39], highlighting the potential of the suggested hybrid approach.

We now propose the following approach based on the ILST for inhomogeneous systems. First, we train a DNN which approximates a mapping between wall-fluid and fluid-fluid forces acting on simple LJ liquids in confinement and its potential parameters. Next, we extract the wall-fluid and fluid-fluid forces acting on the Centre-Of-Mass (COM) mapped trajectories of multiatomic liquids and using these as inputs to the well-trained DNN model, infer the single site coarse-grained potential parameters of the CG bead. The DNN obtained potentials are then used to initialize the RE framework. Using the DNN generated potentials as an initial guess to the RE framework leads to a two-way advantage where both methods gain from each other. First, it extends the applicability of the data-driven approach to include diverse systems with more complex interactions and second, it leads to a critical improvement in convergence of the RE iterations which are slow if used in a stand-alone manner.

The remainder of the paper is organized as follows. First, the inverse problem of liquid state and how it leads to the relationship between fluid forces and CG parameters are described in Section 2.1. A reader only interested in the computational aspect of this work may skip Section 2.1. In Section 2.2 we give details of the CG and AA simulations needed to generate the training data and multiatom fingerprints used for coarse-graining. Next in section 2.3, we describe the data-driven approach along with the architecture of the DNN used for coarse-graining. In Section 2.4 we discuss the Relative Entropy minimization method. Lastly, we assess the accuracy of coarse-graining different multiatomic molecules in Section 3.1 and demonstrate the advantage of using DNN with RE in Section 3.2.

## 2. Methods
### 2.1. Inverse Liquid State Theory (ILST)

Henderson's Theorem[40] guarantees uniqueness of the solution to the inverse problem for homogeneous systems, and techniques such as IBI and DeepILST aim to identify such solutions. In contrast, the inverse problem for inhomogeneous systems has received less attention, despite its relevance to coarse-graining in nanochannels. Given the comprehensive theoretical confirmation of the existence and uniqueness of solutions for inhomogeneous systems in Ref. [6], we will now investigate this matter from a computational perspective. The inverse problem for confined systems is defined as follows:

Consider a system of $N$ particles interacting under the potential $V_N$. As described in Ref. [6], the single particle density under the action of an external potential $U(r)$ can be written as,

$$\rho^{(1)}(r) = \sum_{i=1}^{N} \frac{1}{Z_N} \int exp\left[-V_N(r_1, \ldots, r_i, \ldots, r_N) - \sum_{j=1}^{N} U(r_j)\right] dr_1 \ldots \widehat{dr_i} \ldots dr_N \qquad (1)$$

where the integration is carried over all positions except $r_i$, as indicated by the hat. $Z_N$ is the partition function in the canonical ensemble and $k_B T$ is chosen to be equal to one for simplicity.

The inverse problem for an inhomogeneous system deals with the following question: Given an interparticle potential $V_N(r_1, \ldots \ldots, r_N)$ and distribution $\rho^{(1)}(r)$ of a system such that $\int \rho^{(1)}(r) dr = N$ does there exist a corresponding single particle external potential $U(r)$ that gives the equilibrium particle distribution given by Equation (1) and if it exists, is it unique? Using rigorous theoretical arguments, Ref. [6] shows that for a large class of systems such a potential exists and is unique. In the context of this work, $V_N$ is the pairwise fluid-fluid CG potential while $U(r)$ the pairwise wall-fluid CG potential and $\rho^{(1)}(r)$ the AA fluid density profile. For fluids confined between walls, the wall-fluid potential $U(r)$ is caused solely by wall atoms that depends on the wall-fluid interaction parameters and the distance $z$ from the wall. If the wall atoms are approximated as a continuous medium, then it can be shown that the external potential due to a single wall is given by,[4]

$$U(z) = \frac{2}{3}\pi\rho_w \sigma_{wf}^3 \varepsilon_{wf} \left[\frac{2}{15}(\sigma_{wf}/z)^9 - (\sigma_{wf}/z)^3\right] \qquad (2)$$

Here $\rho_w$ represents wall atom density, $\sigma_{wf}$ and $\varepsilon_{wf}$ the wall-fluid Leonard Jones interaction parameters respectively. Equation (2) along with the preceding definition of the inverse problem in confined systems serves as a starting point for a theoretical treatment of the inverse problem in nanochannels and subsequently relates to coarse-graining. However, this definition is not readily useful from a coarse-graining perspective since the definition assumes that the fluid-fluid potential $V_N$ is already known. This is not the case for most coarse-graining applications in nanochannels where one does not know the fluid-fluid potential beforehand and rather needs to infer it from the corresponding AA system.

We thus reformulate the classical inverse problem in such a way to not keep $V_N$ fixed. However, in doing so we now introduce an additional variable resulting in two unknown potentials $V_N$, $U$ and one known structural property $\rho^{(1)}(r)$ of the nanochannel. Consequently, this leads to an underdetermined system. We address this by introducing a second structural property $g^{(2)}(r_1, r_2)$ which represents the pair-distribution function. For a system with the total interaction potential defined by $W_N$ the pair distribution function is defined as,

$$g^{(2)}(r,r') = \frac{1}{\rho^{(1)}(r) \cdot \rho^{(1)}(r') \cdot Z_N} \int \cdots \int exp[-\beta W_N(r,r',r_3,\ldots,r_N)] dr_3 \cdots dr_N \qquad (3)$$

Where $\beta = 1/k_B T$. Alternatively, the above equation can also be written as,

$$g^{(2)}(r,r') = \frac{1}{\rho^{(1)}(r) \cdot \rho^{(1)}(r')} \left\langle \sum_{i=1}^{N} \sum_{j=1}^{N} \delta(r - r_i)\delta(r' - r_j) \right\rangle \qquad (4)$$

Equation (4) can be used to calculate $g^{(2)}(r,r')$ by Molecular Dynamics via a histogram binning procedure. In the most generic sense $g^{(2)}(r,r')$ (doublet correlation) and $\rho^{(1)}(r)$ (singlet correlation) would be functions of six and three variables, respectively. However, in the case of nanochannels, owing to planar symmetry they reduce to

$g^{(2)}(z_1, z_2, R_{12})$ *and* $\rho^{(1)}(z)$ which are three and one variable functions. Here $z_1$ and $z_2$ are the $z$ coordinates and $R_{12}$ is the axial distance between 2 points used to calculate the pair correlation inside the nanochannel.

The inverse problem of liquid state can now be redefined to compute fluid-fluid potential $V_N$ and wall-fluid potential $U(z)$ that matches the corresponding density $\rho^{(1)}(z)$ and pair-distribution function $g^{(2)}(z_1, z_2, R_{12})$. It is interesting to note that if the external potential $U(z) = 0$ (no walls) then the system becomes homogeneous with $\rho^{(1)}(z) = N/V$ and the pair-distribution function becomes radially symmetric, $g^{(2)}(r_1, r_2) = g^{(2)}(|r_1 - r_2|)$. The problem now simplifies to finding the fluid-fluid potential $V_N$ which matches the corresponding radial distribution function (RDF), which is the well-established inverse problem for homogenous systems. Thus, the inverse problem for a bulk system can be viewed as a special case of an inverse problem for an inhomogeneous system.

In statistical mechanics the inverse problem consists of two facets. Firstly, the inquiry into the existence and uniqueness of the solution, and secondly, the question of whether the solution can be obtained, if it does exist. Although the latter is more relevant and intriguing from a computational perspective, the former warrants some deliberation as well.

The uniqueness of the solution to the inverse problem has been a topic of extensive debate. For instance, the proof of Henderson's Theorem[40] has been shown to incorrectly assume the Gibbs Variational Principle to hold in the thermodynamic limit as demonstrated by Ref. [41]. Additionally, the sensitivity and stability of these solutions have remained problematic, thus limiting their practical applicability. For instance, Ref. [42] demonstrated how vastly different potentials can result in similar-looking radial distribution functions (RDFs). Despite their paramount importance in offering valuable theoretical insights into the physics of the problem, it is imperative to exercise caution when applying such uniqueness results to practical systems.

While we don't establish the uniqueness of the obtained potential, we prioritize the physical consistency of the coarse-grained potentials first by constraining them to a set of physically derived potentials and second by verifying their ability to replicate key structural correlations in the nanochannel.

Although we define $g^{(2)}(z_1, z_2, R_{12})$ as one of the target structural properties to replicate, computing it using MD presents some difficulty as it needs very long simulations compared to the bulk. This is not desirable in the context of ML where the generated training data includes thousands of different systems (as will be elaborated in later sections) and computing the pair correlation for each of these systems entails substantial computational cost. An intuitive theoretical treatment describing the significance and issues pertinent to calculation of pair-distribution in inhomogeneous systems can be found in Ref. [43-45]. Given the intricacy of determining the pair-correlation function, we adopt a simplification based on the observation that, for simple liquids inside a nanochannel, only the density undergoes substantial changes, while the local arrangement of atoms parallel to the axis of the nanochannel remains relatively constant.[46] We corroborate this hypothesis in Appendix A.1 of the supporting material.

This local structure parallel to the axis of nanochannel can be computed as a function of 2 variables by setting $z_1 = z_2 = z$ so that $g^{(2)}(z_1 = z, z_2 = z, R_{12} = R) = g_{\parallel}(z, R)$. We refer to $g_{\parallel}(z, R)$ as the parallel RDF and use it as a metric to test our results as shown in the later sections. Appendix 1 describes the numerical details of calculating $g_{\parallel}(z, R)$ and its value at various locations along the width. It can be seen in supporting information Figure S1 that for a simple liquid (like $CH_4$ and $H_2S$) the parallel RDF hardly changes along the width or confining direction. Having defined the target quantities, we now propose a methodology to compute the coarse-grained wall-fluid and fluid-fluid potentials such that they preserve the structural correlations. Consider the following equation relating density to wall-fluid and fluid-fluid forces,

$$\rho^{(1)}(z) - \rho_0 = e^{\int \beta \left( \bar{f}(z)_{wf} + \bar{f}(z)_{ff} \right) dz} \qquad (5)$$

where $\rho_0$ is the density at the center of the nanochannel and defines the state of the system along with temperature which is fixed at 300 K. Equation (5) shows that the variation in density can be represented as a sum of two force interactions - $\bar{f}(z)_{wf}$ which is the mean wall-fluid force and $\bar{f}(z)_{ff}$ is the mean fluid-fluid force. Furthermore, the mean wall-fluid and fluid-fluid forces can be written as,[45]

$$\bar{f}(z)_{ff} = -\int \rho^{(1)}(z) \, g(z', z, R) \frac{\partial V(z', z, R)}{\partial z} \, dR dz' \qquad (6)$$

$$\bar{f}(z)_{wf} = -\frac{\partial U}{\partial z} \qquad (7)$$

Here $V(z', z, R)$ is the fluid-fluid potential between two fluid particles located at coordinate ($z$, 0) and ($z', R$). Here $z$, $z'$ are the coordinate directions along the width and $R$ along the axis of the nanochannel, respectively. The details of Equation (5) and (6) are given in Appendix A.2 and A.3 of the supporting information. Equations (5-7) elucidate the connection between the mean wall-fluid/fluid-fluid forces and the corresponding interatomic potentials $U$ and $V$, respectively. These equations have traditionally been presented in the context of the anisotropic integral theory [47] and have been extensively applied to obtain density and pair correlations using certain closure equations. While this approach has yielded valuable insights into the structure of confined liquids, it is restricted by the approximate analytical form of closure relations that are only effective for specific types of liquids. We address this issue by using a data-driven approach which learns this non-trivial relationship between forces and interatomic potentials. We define a mapping $\phi$ as follows,

$$\{V, U\} = \phi(\bar{f}(z)_{ff}, \bar{f}(z)_{wf}, \rho_0) \tag{8}$$

The temperature, which is fixed at 300 K, along with the density $\rho_0$ define the thermodynamic state of the system. $\rho_0$ implicitly enters the mapping function because of the boundary conditions on Equation (5). Equation (8) represents a complex mapping between the wall-fluid, fluid-fluid forces, and the corresponding interaction potentials. This mapping is learnt in a data-driven manner using a Feed-Forward DNN which are known to be universal function approximators.[48] It is also worth noting that although $\bar{f}(z)_{ff}$, $\bar{f}(z)_{wf}$ are singlet correlations, the mapping $\phi$ implicitly contains information about the pair correlations which appear in equation (6). The mapping $\phi$ can thus be seen as some transformation over the pair-correlation which is learnt through training the DNN via loss function minimization as will be discussed later. Since we constrain the interaction type to Lennard-Jones potential, the Left-Hand Side (LHS) in equation (8) can be re-written as,

$$\{\sigma_{ff}, \varepsilon_{ff}, \sigma_{wf}, \varepsilon_{wf}\} = \phi(\bar{f}(z)_{ff}, \bar{f}(z)_{wf}, \rho_0) \tag{9}$$

where σ and ε represent the LJ interaction parameters. Once the function '$\phi$' is learnt, Equation (9) can be used to obtain the coarse-grained parameters by using forces obtained from the Center of Mass (COM) mapped trajectories of the All-Atom systems.

**2.2. Data Generation**

In this study, the dataset for the DNN is generated using a MD simulation with 7000 systems of different Lennard-Jones (LJ) fluids confined in a slit-like graphene nanochannel with a fixed width of 8 nm. The lateral dimensions of the graphene layer are $5.0 \times 5.0$ nm². The bulk density values, and force field parameters of the MD simulations are uniformly sampled from the range provided in Table 1. As described earlier, the interatomic interactions are modeled using the 12-6 Lennard-Jones potential with functional form given by:

$$V_{LJ} = 4\varepsilon\left[\left(\frac{\sigma}{r}\right)^{12} - \left(\frac{\sigma}{r}\right)^{6}\right] \tag{10}$$

Here σ, ε are the length and energy scale parameters and $r$ represents the distance between interacting atoms.

**Table 1. Range of values for density and LJ parameters.**

|  | Thermodynamic State | | Parameter Range | |
| --- | --- | --- | --- | --- |
|  | ρ (nm$^{-3}$) | T (K) | σ (Å) | ε (kcal/mol) |
| **Min** | 8.0 | 300 | 1.0 | 0.01 |
| **Max** | 12.0 | 300 | 5.5 | 0.7 |

We perform MD simulations using the large-scale atomic/molecular massively parallel simulator (LAMMPS)[49] with a time step of 1 fs in the NVT (canonical) ensemble with $T = 300$ K. The temperature of the LJ fluid is maintained using the Nosé–Hoover thermostat.[50] Periodic boundary conditions are applied in the x and y directions. In the z direction, we insert an extra vacuum space of three times the actual z dimension to eliminate the slab effect. Simulations are performed with fixed position of graphene walls and each MD simulation is run for 2 ns. The wall-fluid and fluid-fluid forces are calculated on the fly by histogram binning along the width of the nanochannel.

We also evaluate the performance of the DNN in predicting the CG potentials for multiatom molecules including $N_2$, CO, $CH_4$ and $H_2S$. We choose these molecules based on their differences in dipole-moments, moment of inertia, bond-length, mass and number of atoms.[19]

Interactions between multiatom molecules involve both non-bonded and bonded interactions. Force fields for bonded interactions (bonds and angles) are obtained from the GROMOS force-field available in the ATB Repository.[51] The non-bonded interactions can include both short-range van der Waals and long-range electrostatic interactions. The van der Waals interactions are modelled using LJ potentials. The electrostatic interactions are given by: $V_{coulomb(r)} = \frac{q_i q_j}{4\pi\varepsilon_0 r_{ij}}$, where $q_i$, $q_j$ are point charges on interacting atoms $i, j$ respectively, $r_{ij}$ is the distance between them and $\varepsilon_0$ is the dielectric permittivity of vacuum. In MD simulations, the long-range electrostatic interactions are modeled using the particle mesh Ewald algorithm.[52] Simulations are performed using LAMMPS package with a step size of 1 fs. We run the simulations for 2 ns and the trajectory of the atoms are saved for calculating the force and COM profiles. This gives us the COM mapped wall-fluid and fluid-fluid mean force profiles along the width of the channel, which are then used as inputs to the well-trained DNN, which takes this information and relates it to the interaction parameters according to Equation (9).

## 2.3. Neural Network Architecture

A Feed-Forward Deep Neural Network consists of layers of artificial neurons that are connected to each other in a specific architecture. The input layer receives the data, which is then processed and transmitted through hidden layers, which finally produces a transformed value at the output layer. Figure 1 shows the architecture of the DNN used.

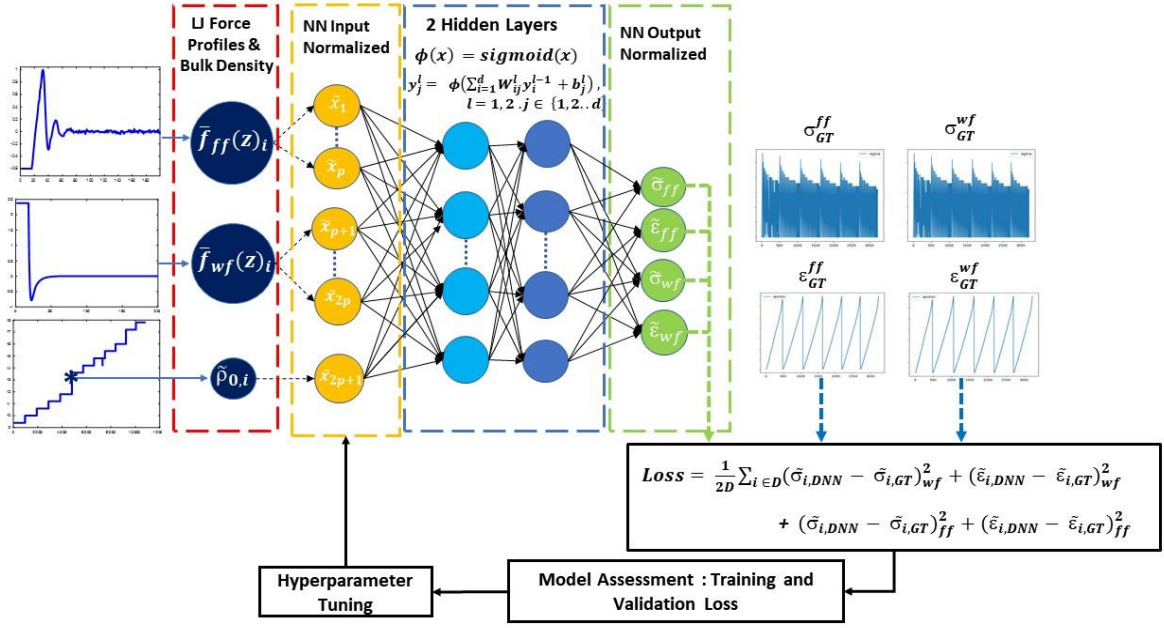

**Figure 1**: The neural network encompasses three inputs: wall-fluid force, fluid-fluid force, and density at the channel center. It generates four outputs, which correspond to the wall-fluid and fluid-fluid LJ parameters. The accuracy of the DNN is evaluated on the validation set after being trained on a set of hyperparameters. The hyperparameters are then fine-tuned to achieve optimal accuracy on the validation set.

The input to the neural network is: $x_i = \left( \bar{f}_{wf}(z)_i,\ \bar{f}_{ff}(z)_i,\ \rho_{0,i} \right)$  (11)

where $i \in D$, the training dataset. $x_i$ is the input vector constructed by concatenating the wall-fluid forces, fluid-fluid forces which are arrays of dimension $p$ each and density value at the center of the channel. The number of bins $p$ along the width of nanochannel, was chosen to be 200. The choice of $p$ which affects the input dimensionality was influenced by two factors. Firstly, it was noted that a small bin size led to noisy histogram force profiles, while a large bin size caused "smearing" of force profiles and thus loss of important features. The second consideration was the effect of feature size on the training of the DNN. A large dimensionality of input vector typically increases the complexity of learning task for the DNN, thereby requiring more parameterization. This was also observed during training where both the number of neurons and the network's depth needed to be increased to minimize the loss function as the bin size decreased. This is believed to be a result of both noise in the input profiles and the high dimensionality of input features. Nevertheless, the value of $p$ was chosen to keep the input dimensionality as low as possible without causing undue "smearing" of the histogram profiles.

Each neuron in a DNN processes the input data by applying a non-linear activation function to a linear combination of the input data and a set of weights and biases. This transformation is given by: $\phi_k(x) = \phi_k(W_k \phi_{k-1}(x) + b_k)$, where $\phi_k$ is the activation function in the $k^{th}$ layer. $W_k$, $b_k$ are the set of weights and biases in the $k^{th}$ layer, respectively. Commonly used activation functions include sigmoid, ReLU and tanh. The activation function is used to introduce non-linearity into the network, allowing the DNN to learn more complex functions and representations. We found that using a sigmoid activation function for all the layers gave the best results. The parameters $W_k$ and $b_k$ are then adjusted during training to minimize the loss given by,

$$Loss = \frac{1}{2D} \sum_{i \in D} (\tilde{\sigma}_{DNN,i} - \tilde{\sigma}_{GT,i})^2_{wf} + (\tilde{\varepsilon}_{DNN,i} - \tilde{\varepsilon}_{GT,i})^2_{wf} + (\tilde{\sigma}_{DNN,i} - \tilde{\sigma}_{GT,i})^2_{ff} +$$
$$(\tilde{\varepsilon}_{DNN,i} - \tilde{\varepsilon}_{GT,i})^2_{ff} \qquad (12)$$

where the subscripts $'DNN', 'GT'$ represent the predicted and ground-truth values, respectively. Tilda represents normalized values for all the output parameters. The loss minimization is done using gradient descent techniques. We got the best performance using the Adam algorithm[53] with a learning rate of 0.1 and a batch size of 64.

An important aspect of ML is to make the model predict data outside the training dataset and prevent overfitting. Several other techniques like dropout, ensemble averaging, and batch normalization have been widely used to deal with this issue.[54-55] Here, we use dropout in the second layer which drops certain nodes randomly with a rate of 0.3 thus forcing the network to learn redundant features and preventing it from depending too much on any individual neuron. When training a neural network, it is also important to use different sets of data for training, validation, and testing. We randomly separate the dataset into three groups for training (70% of the total dataset), validation (15%) and testing (15%). The training data is used to adjust and update weights and biases of the neural network during the training process by calculating gradients using backpropagation. The validation data is used to evaluate the performance of the model during the training process and to tune the hyperparameters of the model, such as the learning rate and the number of hidden layers. We got the best performance on the validation dataset using 2 layers with 128 and 64 nodes, respectively. Once a good performance has been achieved on both training and validation dataset, the final performance is evaluated on the testing dataset. The error on the testing data gives an estimate of how well the model is likely to perform on unseen data.

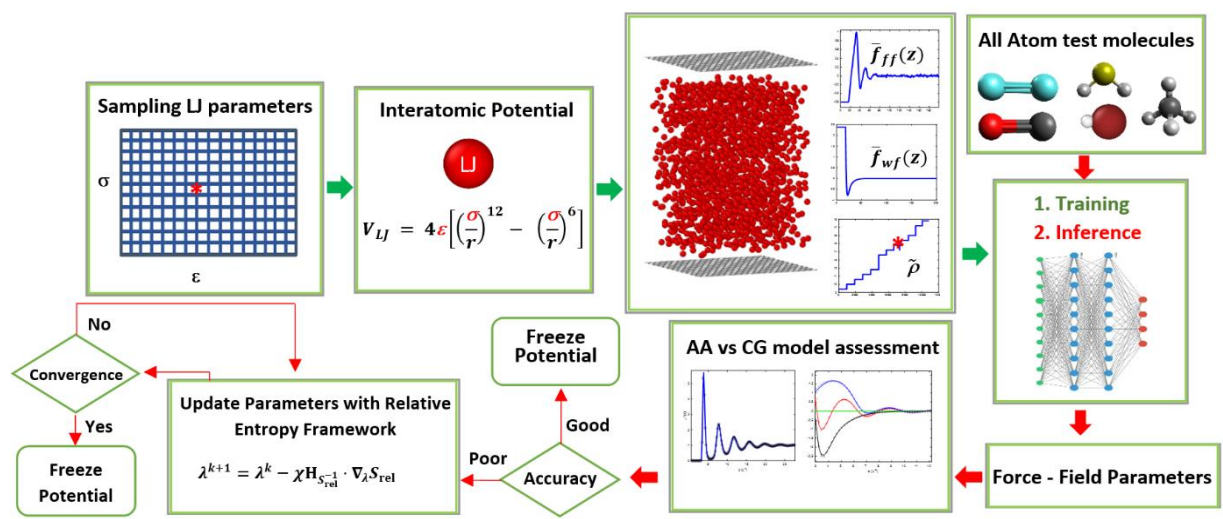

**Figure 2**: Methodology employed to coarse-grain simple multiatomic molecules in confinement. First MD simulations are performed for a range of LJ fluids at different densities to generate force maps which are used for training, testing and validation of the DNN (green arrows). Once the DNN is well-trained we carry out inference where force maps from the

Center of Mass mapped All-Atom trajectories are used as inputs to the DNN and the predicted outputs of the DNN serve as the coarse-grained LJ parameters (red arrows). The density and force variation of the coarse-grained systems are then compared with their All-Atom counterparts. If the accuracy is low the potential is further fine-tuned by performing Relative Entropy updates.

Figure 2 describes this sequential approach employed for obtaining coarse-grained wall-fluid and fluid-fluid potentials. The framework consists of 2 distinct parts i) Training (highlighted in green) and ii) Inference (highlighted in red).

During the training phase, the DNN learns the non-trivial relationship between the mean-force profiles and the corresponding LJ force fields for simple LJ liquids in confinement as given by Equation 9. Subsequently, this well-trained DNN is repurposed to perform coarse-graining on more intricate multiatomic liquids, a process referred to as transfer learning. In this context, transfer learning entails utilizing a DNN originally trained on simple LJ liquids to perform coarse-graining on complex multiatomic liquids. To achieve this, we initially compute force profiles for the wall-fluid and fluid-fluid interactions along the trajectories of the mapped Center of Mass for multiatomic molecules. These force profiles are then employed to conduct inference with the adeptly trained DNN. The DNN's output yields the coarse-grained parameters corresponding to the multiatomic molecule under consideration. Subsequently, we gauge the efficacy of transfer learning by comparing the results of the coarse-grained simulations, denoted as CGNN, with their All-Atom counterparts. As mentioned earlier, the LJ potential form although physically derived is less accurate for certain liquids where electrostatic or multibody correlations are dominant. Using the DNN derived LJ potentials as a starting point more flexible numerical potentials with optimal functional form are derived using Relative Entropy Minimization.

## 2.4. Relative Entropy Minimization

The measure known as Relative Entropy, which is derived from information theory, calculates the level of similarity between two probability distribution functions,[56] and can be expressed as follows:

$$S_{rel} = \sum_i p_{AA}(i) ln\left(\frac{p_{AA}(i)}{p_{CG}(M(i))}\right) + \langle S_{map}\rangle_{AA} \qquad (13)$$

where $S_{rel}$ is the Relative Entropy and $i$ corresponds to a particular set of atomic positions in the AA ensemble. $M$ is the AA to CG mapping operator and $S_{map}$ is the mapping entropy due to all states in AA system that map to the same CG configuration. It can be shown that $S_{map}$ doesn't depend on the CG potential and is rather a function of the mapping operator $M$. In this case the mapping $M$ assigns CG positions to the COM of the molecules in the AA ensemble. $p_{AA}$ and $p_{CG}$ are phase space probabilities for the AA and CG system corresponding to configuration $i$ which depend on the interatomic potentials. RE can thus be used as a metric to maximize the overlap between AA and CG configurations. One major drawback of this method is that it converges slowly for an arbitrary initial guess of parameters. A recommended[57-58] way of initializing the parameters for RE is to invert the RDF according to the equation, $-k_B T \ln g(r) = v(r)$, where $g(r)$ is the RDF and $v(r)$ is the interparticle potential, which we refer as the "Boltzmann Inverted" potential guess. Although physically consistent in some sense, this equation holds only for homogeneous systems and in the limit where density of the system approaches zero.

In this work, we try to address these limitations by exploiting the similarity between force-based methods and Relative Entropy. In the past, many works have explored the relation between Relative Entropy and force-based methods[59]. Although they are based on different mathematical formulations and physical principles, under certain conditions it is shown that they converge to similar solutions.[60] Given this similarity we explore the possibility of using the DNN generated potentials as an initial guess for the RE framework to improve its convergence.

## 3. Results and Discussion

### 3.1. Performance of DNN for LJ fluid and coarse-graining simple liquids

The DNN model is trained using the LJ fluid dataset generated in Section 2.2 and optimized via minimizing the loss function in Equation (12). We first assess the performance of the DNN model by comparing the predicted LJ potential parameters with their corresponding ground truth values. To quantify the training and testing error, we use the weighted mean absolute percentage error (MAPE) metric defined as $\epsilon_{\text{MAPE},j} = 100 \times \frac{\sum_{i \in D} \left| v_{j,\text{DNN}}^{(i)} - v_{j,\text{GT}}^{(i)} \right|}{\sum_{i \in D} \left| v_{j,\text{GT}}^{(i)} \right|}$, where $v$ represents one of the four LJ potential parameters, $\sigma_{ff}, \varepsilon_{ff}, \sigma_{wf}, \varepsilon_{wf}$. $i$ represents the $i^{th}$ data point and $j = wf \text{ or } ff$ corresponding to the wall-fluid and fluid-fluid potential parameters. Figure 3(a)-3(d) show the one-to-one comparison of ground truth values and DNN predicted LJ potential parameters for the training dataset. We can see that the DNN model is well-trained to approximate the mapping (see Eq 9) with maximum MAPE $\leq$ 5.4 %. To check model generalization, the DNN model is then tested using the testing data which is held out and not seen during training. The testing results are shown in Figure 3(e)-3(h). We observe that most points in testing dataset lie close to the ideal line with the maximum MAPE $\leq$ 2.0 %.

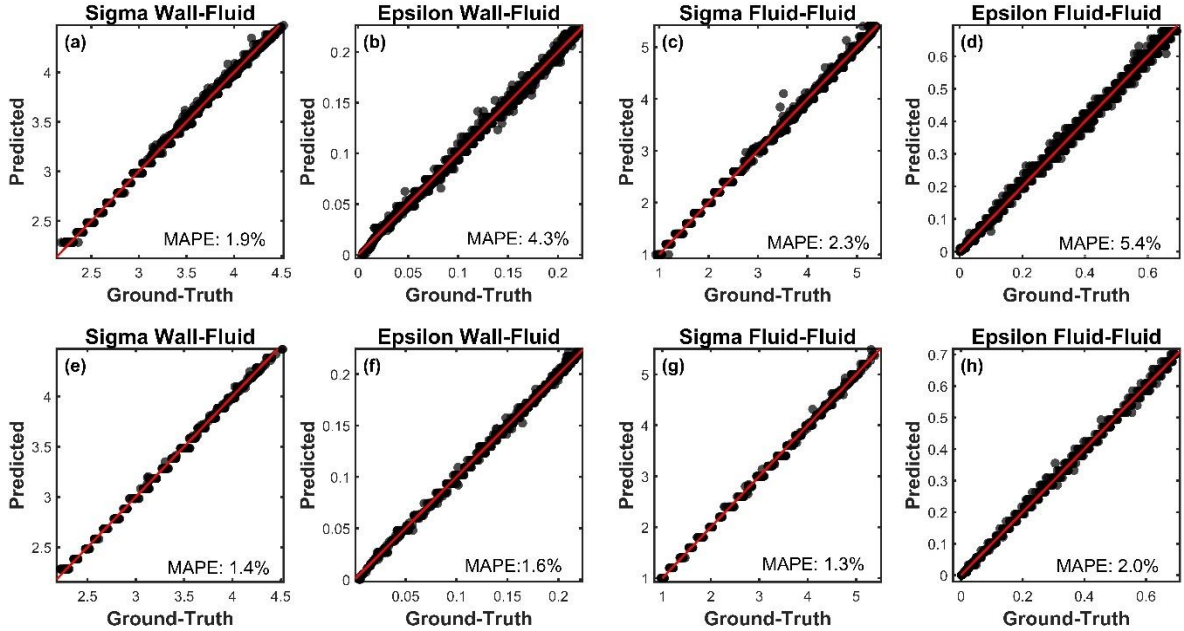

**Figure 3**: Comparison of the predicted and ground-truth LJ parameters for (a - d) Training and (e-h) Testing data.

The DNN model can thus predict the LJ potential parameters for unseen data. Next, we transfer the DNN knowledge of LJ fluids to predict the coarse-grained force field parameters for simple multiatomic molecules in confinement. In the coarse-graining process, a multiatom molecule is coarse-grained into a single bead such that the COM mapped wall-fluid and fluid-fluid interactions are preserved in the single bead system. As shown in figure 2 (red arrows in the figure), the wall-fluid and fluid-fluid force profiles obtained from the AAMD simulations and the corresponding density $\rho_0$ are fed as inputs to the well-trained DNN model. Using this information, the DNN model predicts the coarse-grained force field parameters used to represent multiatomic molecules as single beads within the nanochannel.

In order to evaluate the accuracy of the proposed method for coarse-graining, the CG force field parameters were utilized to carry out molecular dynamics (MD) simulations on single bead systems. These simulations, referred to as CGNN simulations, are performed with the same LAMMPS settings described in Section 2.2. A comparison is then made between the structural properties and force profiles of AA systems and those obtained from the CGNN simulations. The discrepancy between AA and CG profiles is measured using the error given as $\% \, err_{COM/RDF} = \frac{\int_0^r |X_{CG}(r)-X_{AA}(r)|r^2 \mathrm{d}r}{\int_0^r X_{AA}(r)r^2 \mathrm{d}r}$ and $\% \, err_{Force} = \% \, err = \frac{|f_{CG}(r)-f_{AA}(r)|\mathrm{d}r}{|min(F_{AA}(r))|}$

Here $f$ is the force, $X$ denotes either the Center of Mass (COM) density or the parallel RDF. These quantities of the corresponding AA or CG system are calculated at the $i^{th}$ bin and $N_b$ denotes total number of bins. Figure 4 shows results for the COM profiles for both AAMD and CGNN simulations of $N_2, CO, CH_4,$ and $H_2S$. The COM profiles obtained from DNN parameterized CGNN are in good agreement with AAMD results for $H_2S$, $N_2, CO$ and $CH_4$ with error $err_{COM} \in [5.42\%, 2.97\%, 4.04\%, 2.40\%]$, respectively.

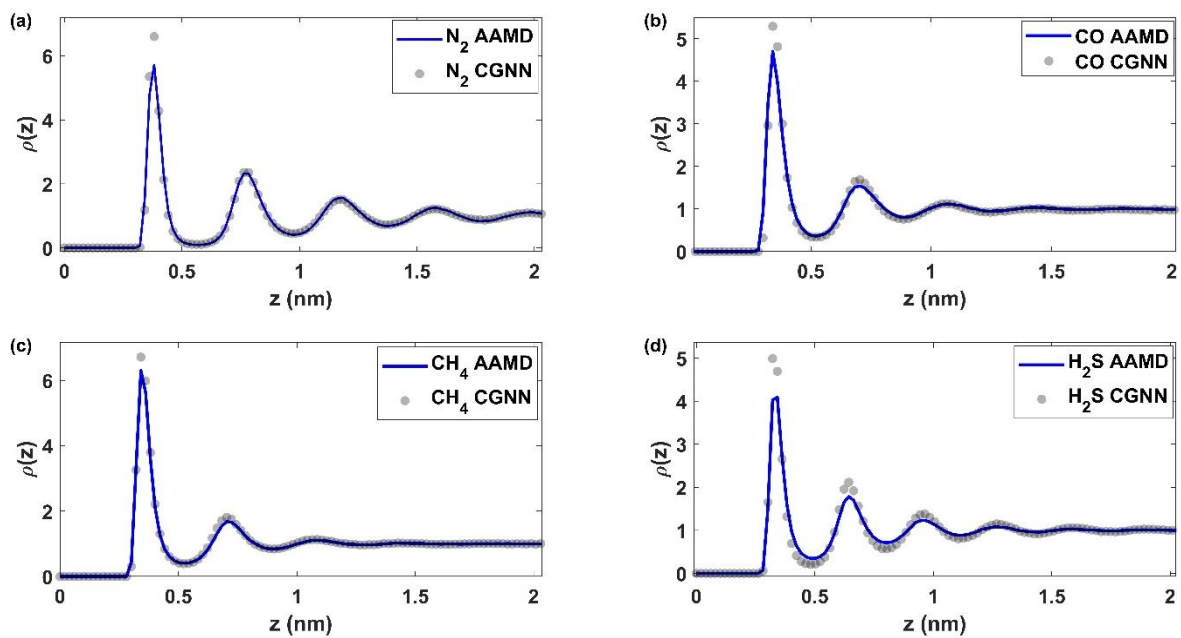

**Figure 4**: Comparison of density profiles of AA and CGNN systems of different simple multiatomic liquids.

Figure 5 compares the parallel RDF, which is the second structural property we seek to replicate. The errors for $H_2S$, $N_2$, CO and $CH_4$ are $err_{RDF} = [2.51\%, 1.92\%, 1.12\%, 0.93\%]$, respectively. In addition to the fluid structure, the total molecular force in the AAMD simulations is also preserved in the CGNN simulations for the above four types of molecules. It can be seen in Figure 6 that the total molecular force obtained from CGNN simulations agrees well with the AAMD results for $H_2S$, $N_2$, CO and $CH_4$ with $err_{force} \in [2.32\%, 2.13\%, 1.82\%, 0.84\%]$.

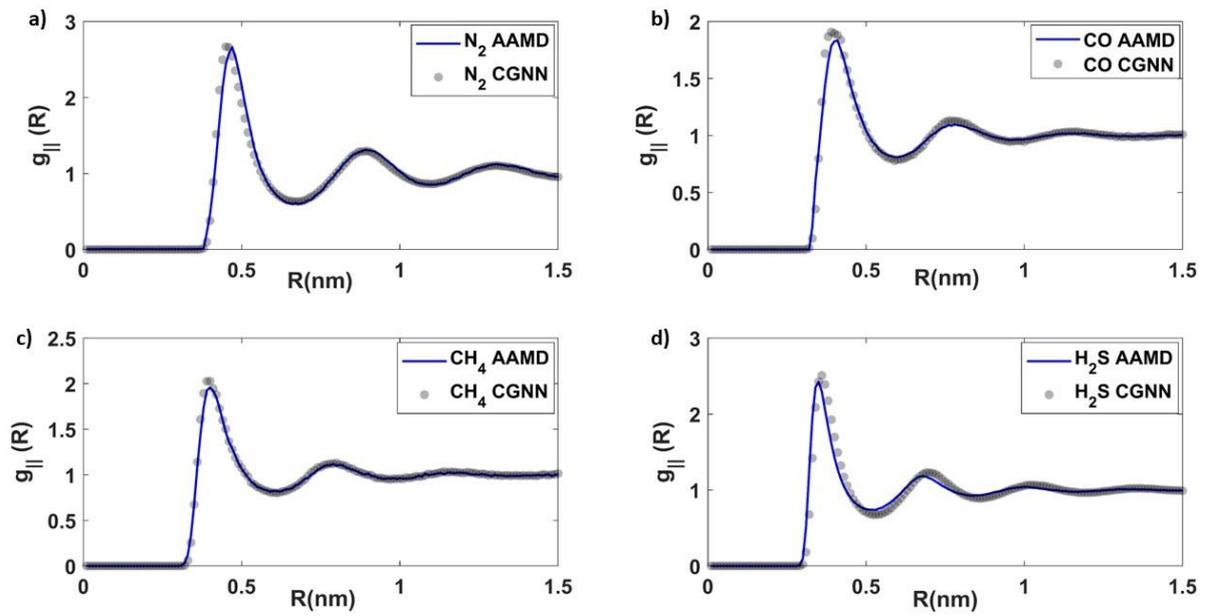

**Figure 5**: Comparison of parallel RDF profiles of AA and CGNN systems of different simple multiatomic liquids.

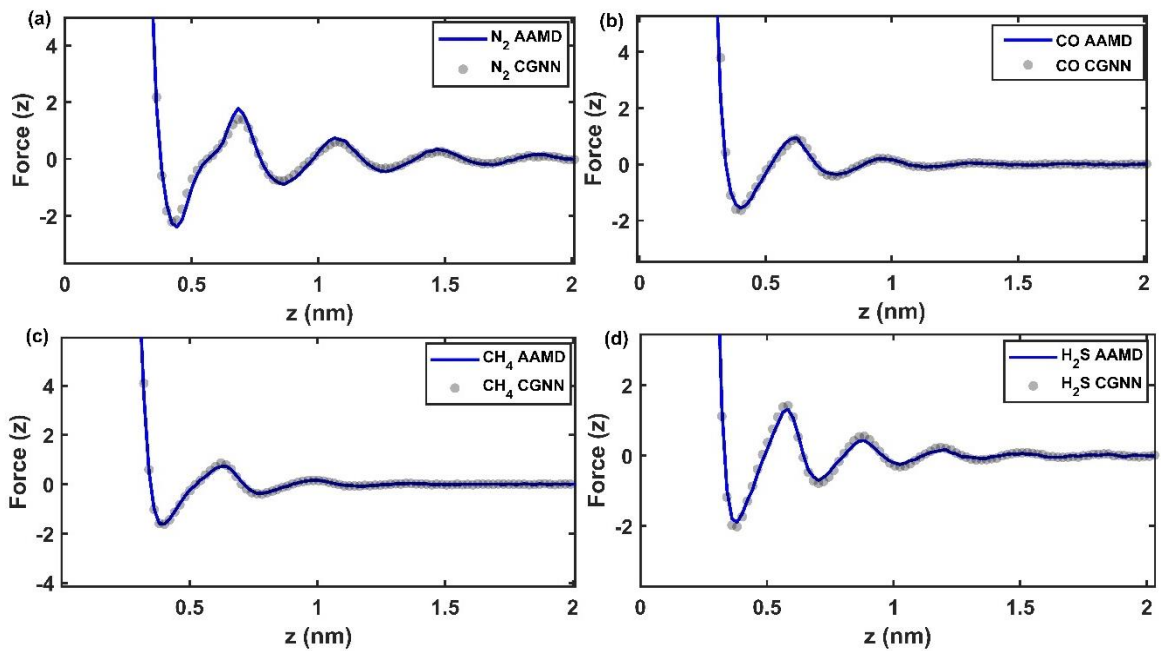

**Figure 6**: Comparison of force profiles of AA and DNN predicted CG systems of different multiatomic simple liquids.

It should be noted that $N_2$ and $CH_4$ are non-polar while CO and $H_2S$ are polar molecules that exhibit long-range electrostatic interactions. Since the training data used by the DNN includes only neutral LJ liquids, the transfer learning process can only distill information of short-ranged van der Waals interactions from the AA force maps and approximates the electrostatic component as a short-range potential. This approach of locally approximating slowly varying attractive forces was originally proposed by Ref. [61] for studying inhomogeneous simple liquids. Their approach demonstrated that an inhomogeneous fluid's structure can be described by an equivalent reference fluid with a short ranged purely repulsive core under the influence of an external field. This field can be calculated using a self-consistent iterative approach based on the BBGKY hierarchy. However, such a truncated model leads to errors as the long-range interactions start becoming significant. For instance, the DNN parameterized CGNN displays a larger deviation on all target properties on H2S, which has a much higher dipole moment (1.84 Debye) compared to CO (0.01 Debye), $N_2$ and $CH_4$ (non-polar).

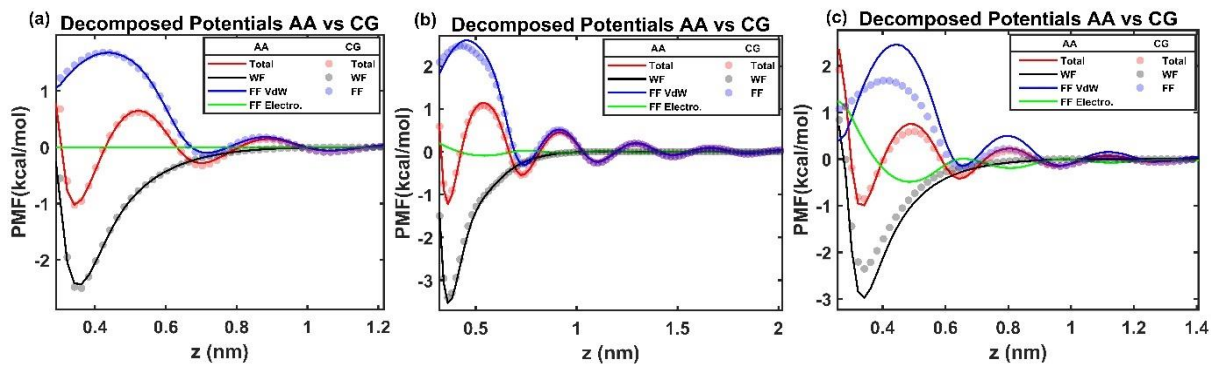

**Figure 7**: Comparison of wall-fluid (WF), fluid-fluid van der Waals (FF VdW), electrostatic (FF Electro) and total potential of mean force (PMF) profiles for (a) CO (b) HBr and (c) $H_2S$

We conduct a quantitative evaluation of the influence of long-range electrostatic forces on the efficacy of transfer learning process. To effectively assess the electrostatic potential contribution, we first decompose the total PMF into wall-fluid potential and fluid-fluid potential. The fluid-fluid PMF can further be decomposed into van der Waals and electrostatic interactions. The electrostatic PMF computed includes both the short-range and the long-range interactions which are computed in real and reciprocal space respectively. Since the carbon atoms in the graphene surface have zero charge in AAMD simulations, only the fluid-fluid interactions cause electrostatic contribution to the total PMF. The decomposed potentials for the 3 types of molecules with varying dipole moments namely, CO (0.01 Debye), HBr (1.59 Debye) and $H_2S$ (1.84 Debye) are shown in Figure 7. The electrostatic contribution to the total PMF is weak for CO and therefore it is observed that all the three decomposed PMF components namely wall-fluid, fluid-fluid, and total potentials from the CGNN simulations are in good agreement with their corresponding AAMD counterparts. Consequently, the DNN predicts the CG parameters to effectively replicate both the fluid structure and molecular force of the target AA system. For HBr, the electrostatic potential has a stronger contribution than CO near the interface. This leads to a slight deviation closer to the wall, although the overall PMF still remains preserved. The discrepancy is greater for $H_2S$, which has a relatively stronger electrostatic contribution. However, it is noteworthy that while the separate contributions vary considerably, the collective coarse-grained (CG) potential of mean force (PMF) remains similar to the All-Atom PMF. This suggests that the DNN approximates the long-range forces as short-range components and utilizes them to deduce the CG parameters which closely matches with the All-Atom PMF.

**3.2. Initializing RE method with DNN potentials as initial guesses**

The preceding section revealed that the CGNN simulation falls short in accurately predicting the COM profile for molecules with relatively stronger electrostatic interactions where the short-ranged LJ potential is not a good approximation. To overcome this limitation of using a Lennard-Jones potential form, we sequentially combine the DNN method with Relative Entropy minimization based coarse-graining method which allows for more flexible numerically optimized potentials. We employ the DNN method to generate an initial set of Lennard-Jones parameters as described in Section 2.3. The resulting LJ functional form is then modelled as a cubic spline whose coefficients now define the shape of the wall-fluid and fluid-fluid potential. This numerical potential then serves as a robust starting point to optimize the RE objective function. This RE framework then iterates over the initial guess and updates the fitting coefficients of the cubic spline thereby shaping the potential profiles in such a way to minimize the Relative Entropy between AA and CG ensembles.

We perform the RE based coarse-graining using the VOTCA package.[16, 60] As shown in Figure 8, the DNN and RE framework provide potentials that are nearly identical for simple non-polar liquids like $CH_4$. As shown in the previous section, in case of $H_2S$ the CGNN predictions for density show a relatively higher error of 5.42%, while other molecules with relatively lower electrostatics show good accuracy with errors less than 5% which we use as a threshold cutoff criterion. The RE framework then further refines the $H_2S$ potentials predicted by the DNN model, resulting in a CG system whose density closely matches the AA system's density, as illustrated in figure 9(c)-9(d).

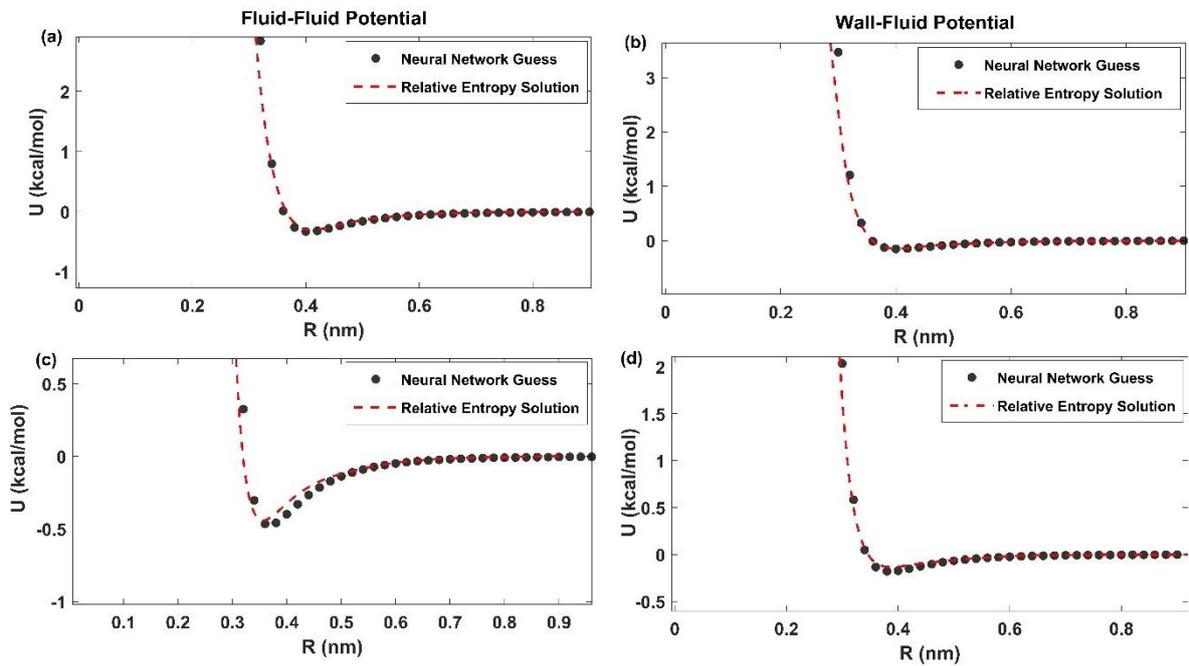

**Figure 8**: Comparison of DNN predicted initial guesses and converged RE solution for (a, b) CH4 and (c, d) $H_2S$.

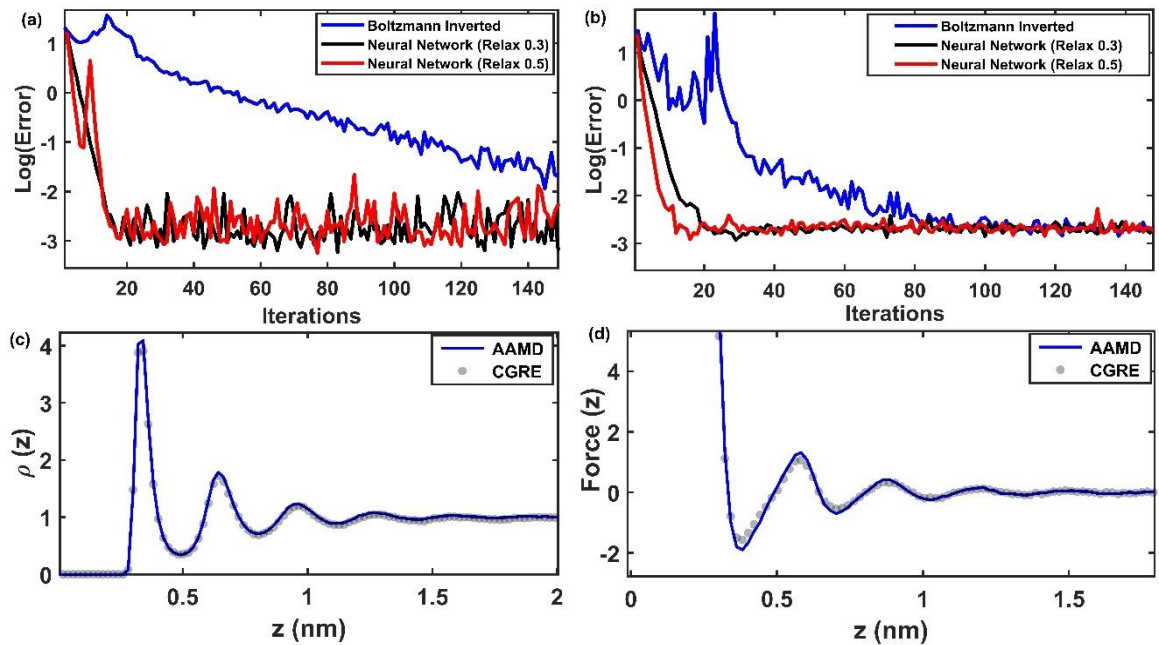

**Figure 9**: (Top) Convergence characteristics for different initialization schemes. (Bottom) Density/Force profile comparison for $H_2S$ with potentials obtained from RE Minimization procedure initialized with CGNN potentials. "Relax" here refers to relaxation factors used for RE update.

From Figure 9(a)-(b) it is seen that initialization has a significant effect on convergence for the RE method. Our findings indicate that a poor initial guess can cause the RE iterations to diverge resulting in termination of the iterative process. We observe that for both $CH_4$ and $H_2S$, using the Boltzmann Inverted (BI) potential as the initial guess leads to a slow or weak convergence resulting in a significant computational cost. While significant modifications can be made to improve the BI convergence using different pre-factors and relaxation settings, the process remains slow and ad-hoc. In contrast, using the DNN generated LJ potential as the initial guess results in fast convergence, as demonstrated in Figure 9(a, b). The results indicate that, with a robust initial guess generated by the DNN model, the RE framework converges within 25 iterations for both $H_2S$ and $CH_4$. In comparison, when using a Boltzmann Inverted initial guess (with carefully tuned convergence parameters), the RE framework requires at least 100 iterations for $H_2S$ and even more for $CH_4$ to converge. Since each iteration of the RE framework involves a separate CG simulation, this improvement translates to a reduction of 100 CG simulations compared to the Boltzmann Inverted initial guess thus demonstrating a significant improvement in convergence and computational cost.

## 4. Conclusions

In this study, we have explored the application of inverse liquid state theory in the purview of confined systems and in doing so identified key correlations that characterize the structure of inhomogeneous fluid. More importantly, the solution to the inverse liquid state problem for inhomogeneous system is seen as a route to coarse-graining simple multiatomic molecules where the identified structural correlations of the All-Atom system are to be preserved. The wall-fluid and fluid-fluid forces are shown to contain necessary information required to preserve structural properties of the AA system. We then develop a novel data-driven approach where a DNN approximates the complex mapping between wall-fluid/fluid-fluid forces of simple fluids and corresponding potential parameters. The well-trained DNN is then used to coarse-grain simple multiatomic molecules by utilizing transfer learning. Our model accurately reproduces density, parallel RDF and force profiles across the channel width for various molecules at different densities particularly when the electrostatic interactions are not dominant.

However, we found that this limitation inherent to transfer learning can be addressed by using the DNN framework in conjunction with Relative Entropy minimization based coarse-graining method. We discovered that the DNN generated potential provides an excellent initial guess for the RE framework, leading to significant improvements in convergence. Furthermore, we demonstrated that the combination of data-driven approach with the RE method can help fine-tune the CG potential obtained using the DNN. This is evidenced by the improved agreement between the AA and CG density profiles for $H_2S$, which were achieved through RE based iterative refinement of the initial DNN potentials. Overall, the synergistic combination of data-driven and physics-based Bottom-Up approaches results in improved fidelity and convergence, allowing for more efficient and more reliable coarse-graining of complex molecular systems.

ASSOCIATED CONTENT

**Supporting Information**

Calculation of parallel RDF and details of Equations (5,6). (PDF)

Results and supporting scripts used for coarse-graining procedure in this work can be accessed at https://github.com/ishannadkarni1997/CONF_ILST.

AUTHOR INFORMATION:

**Corresponding Author**

*Email - aluru@utexas.edu

**Notes**

The authors declare no competing interests.

ACKNOWLEDGEMENTS:

The work on deep learning was supported by the Center for Enhanced Nanofluidic Transport (CENT), an Energy Frontier Research Center funded by the U.S. Department of Energy, Office of Science, Basic Energy Sciences (Award No. DE-SC0019112). All other aspects of this work were supported by the National Science Foundation under (Grant Nos. 2140225 and 2137157). The authors acknowledge the Texas Advanced Computing Center (TACC) at The University of Texas at Austin for providing access to the Lonestar6 resource that has contributed to the research results reported within this paper. We also acknowledge the use of the Extreme Science and Engineering Discovery Environment (XSEDE) Stampede2 at the Texas Advanced Computing Centre through Allocation No. TG-CDA100010.

**For Table of Contents Only**

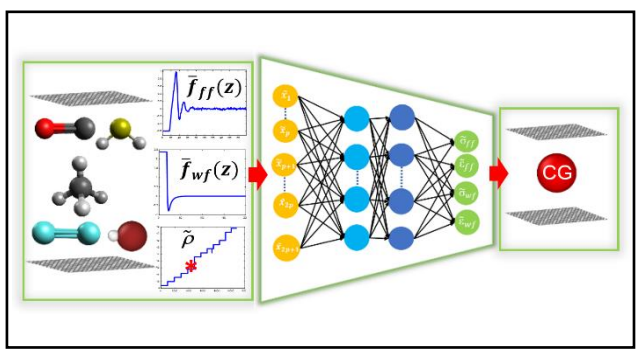